\documentclass[12pt]{article}
\usepackage{amsfonts}
\usepackage{bbding}
\usepackage{mathrsfs}
\usepackage{amssymb}
\usepackage{amsmath}
\usepackage{graphicx}
\usepackage{color}
\usepackage{epsfig}

\newcommand{\calS}{{\cal S}}

\newcommand{\calF}{{\cal F}}

\newcommand{\etal}{{\it et al\/}.\ }

\newcommand{\dbz}{{d{\mbox{\boldmath $Z$}}}}

\def\p{\partial}

\def\a{\alpha}
\def\b{\beta}

\def\D{\Delta}
\def\e{\epsilon}
\def\g{\gamma}

\def\o{\omega}
\def\s{\sigma}
\def\Sg{\Sigma}
\def\td{\tilde}

\def\defn{\stackrel{\triangle}{=}}

\def\td #1{{\tilde #1}}

\begin{document}
\title{\bf Inflation-rate Derivatives: From Market Model to Foreign Currency Analogy}
\author{
Lixin Wu\thanks{ 
Part of the results have already been published in I want to thank participants of Financial Mathematics Seminar in Peking University in December 18, 2007, and BFS 2008 Congress, London, for their comments. 
All errors are ours. Email: malwu@ust.hk.}  \\
Department of Mathematics\\
University of Science and Technology \\
Clear Water Bay, Kowloon \\
Hong Kong
 \\
 \\
First version: {August 1, 2008} \\
This version: {January 4, 2010}
\date{}}
\maketitle

\newpage
\begin{abstract}
In this paper, we establish a market model
for the term structure of forward inflation rates based on the risk-neutral dynamics of nominal and real zero-coupon bonds. Under the market model, we can price inflation caplets as well as inflation swaptions with a formula similar to the Black's formula, thus justify the current market practice.
We demonstrate how to further extend the market model to cope with volatility smiles. Moreover, we establish a
consistency condition on the volatility of real zero-coupon bonds using arbitrage arguments, and with that re-derive the model of Jarrow and Yildirim (2003) with real forward rates based on ``foreign currency analogy", and thus interconnect the two modeling paradigms.
\end{abstract}

\vspace*{0.2in}



\newpage

\section{Introduction}
\textcolor{black}{``Foreign currency analogy"} has been the standard technology for
modeling inflation-linked derivatives (Barone and Castagna, 1997; Bezooyen et al. 1997; Hughston, 1998; Jarrow and Yildirim, 2003).
In this approach, real interest rate, defined as the difference
between nominal interest rate and inflation rate, is treated as the
interest rate of a foreign currency, while the consumer price index
(CPI) is treated as the exchange rate between domestic and the
foreign currency. To price inflation derivatives, one needs to
model nominal (domestic) interest rate, foreign (real) interest
rate, and the exchange rate (CPI). A handy solution for modeling
inflation derivatives is to adopt the Heath-Jarrow-Morton's (1992) framework separately for
both interest rates, and bridge them with a lognormal exchange-rate process. For a
comprehensive yet succinct introduction of the pricing model under
the so-called ``HJM foreign currency analogy", we refer readers to Manning and Jones (2003).

Although elegant in theory, a Heath-Jarrow-Morton type model is known to be inconvenient for derivatives pricing. The model takes unobservable instantaneous nominal and real forward rates as state variables, making it hard to be calibrated to most inflation derivatives, as their
payoffs are written on CPI or simple compounding inflation rates. 


Aimed at more convenient pricing and hedging of inflation derivatives, a
number of alternative models have been developed over the years.
These models typically adopt lognormal dynamics
for certain observable inflation-related variables, for examples, CPI index (Belgrade and Benhamou, 2004; Belgrade et al., 2004)
or forward price of real zero-coupon bonds (Kazziha, 1999; Mercurio, 2005). Recently extensions of models along this line have incorporated more sophisticated driving dynamics like stochastic volatility (Mercurio and Moreni, 2006 and 2009) and a jump-diffusion (Hinnerich, 2008).
Besides, there are also papers that address various issues in inflation-rate modeling, like ensuring positive nominal interest rates by Cairns (2000), and estimating inflation risk premiums by Chen et al. (2006), among others.
Although most of
these models achieve closed-form pricing for certain derivatives, they carry
various drawbacks, from complexity of pricing to not being a proper term
structure model that describes the co-evolution of nominal interest rates and
inflation rates. In the meantime, market practitioners have generally adopted a model of their own, the so-called market model based on displaced diffusion dynamics for ``forward inflation rates"\footnote{There exist various version of ``forward inflation rates" in literature.}. The market model of practitioners, however, has not appeared in literature available to public.


In this paper, we put the market model in a rigorous footing. We take, in particular, nominal
zero-coupon bonds and real zero-coupon bonds as model
primitives, define the term structure of forward inflation rates, and rigorously establish the practitioners' market
model, where forward inflation rates follow displaced-diffusion
processes. Such displaced diffusion processes lead naturally to a Black like formula
for inflation caplets, and, after some light approximations, inflation swaptions. Owing to this closed-form formula, the market model can be calibrated to inflation
caps, floors and swaptions using an existing technology for calibrating the LIBOR market model.
For theoretical interests, we also establish a HJM type model for instantaneous inflation forward rates.

There are a number of important results arisen from our research. First, we define forward inflation rates based on arbitrage arguments, which is thus unique and thus should change the situation of the coexistence of multiple ``forward inflation rates" in literature. Second, we establish that the martingale property of forward inflation rates under their own cash-flow measures\footnote{A forward measure with delivery date equal to the maturity date of the forward inflation rate.}. Third and perhaps most importantly, we
discover a so-called {\it consistency condition}, a necessary condition for the absence of arbitrage with the volatilities of nominal and real zero-coupon bonds, and show under this condition that the model we have developed with forward inflation rates is actually consistent with the model developed by Jarrow and Yildirim (2003) with forward real rates, in the sense that we can derive one model from the other. Fourth, the pricing of year-on-year inflation-index swaps becomes model free. Lastly, we have clarified that the volatility of the CPI index should be zero\footnote{Thanks for the comments of an anonymous referee.}, which somehow undermines the notion of ``foreign currency analogy" for inflation derivatives.

The extended market model for inflation rates also serves as a platform for
developing more comprehensive models. For instance, in order to capture volatility smiles
or skews of inflation derivatives, one may adopt stochastic volatilities or jumps
to the driving dynamics, in pretty much the same ways these random dynamics are
incorporated into the standard LIBOR market model.  We refer readers to Brigo and Mercurio (2006) for a
comprehensive introductions of extensions to LIBOR market models.

The rest of the paper is organized as follows. In section 2, we
introduce major inflation derivatives and highlight real
zero-coupon bonds, part of our primitive state variables. In section 3, we define the notion of forward inflation rates and establish an extended market model. We then present pricing formulae of major inflation-rate
derivatives under the extended market model. A
Heath-Jarrow-Morton type model in terms of continuous compounding forward nominal
and inflation rates is also established as a limiting case. Section 4 is devoted to the pricing of inflation-indexed
swaption under the market model, where we produce a closed-form formula for swaption prices. In section 5, we will discuss the comprehensive
calibration of the market model, and demonstrate calibration results with market data. In section 6, we demonstrate the construction of of smile models with in particular the SABR-type extension of the market model. Finally in section 7 we conclude the paper. The proofs of some propositions are put in the appendix.

\section{CPI Index and Inflation Derivatives Market}

Inflation-rate security markets have evolved steadily over the
last decade, with the outstanding notional values growing from about 50
billion dollars in 1997 to over 1 trillion dollars in 2007. There are inflation-linked securities in most major
currencies, including pound,
Canadian dollar, yen and of course, Euro and U.S. dollar.
The global daily turnover on average
exceeded \$3 billions a day in 2007, which is largely dominated by Euro and dollar
denominated securities. Nonetheless, by comparing to the sizes of LIBOR or credit markets, one has to conclude that the interest on inflation securities has been
tepid in the past, but there are encouraging signs
that the situation is changing (Jung, 2008).

The payoff functions of inflation-linked securities depend on
inflation rates, which are defined using Consumer Price
Index (CPI). The CPI represents an average price of a basket of
services and goods, the average price is compiled by official
statistical agencies of central governments. The evolution of CPI indexes in
both Europe and United States are displayed in Figure 1, which show
a trend of steady increase. Since 2008 there has been a concern on the possible
escalation of inflation in the near future.

%

\vspace*{0.2in}
\centerline{\epsfxsize=2.5in \epsfbox{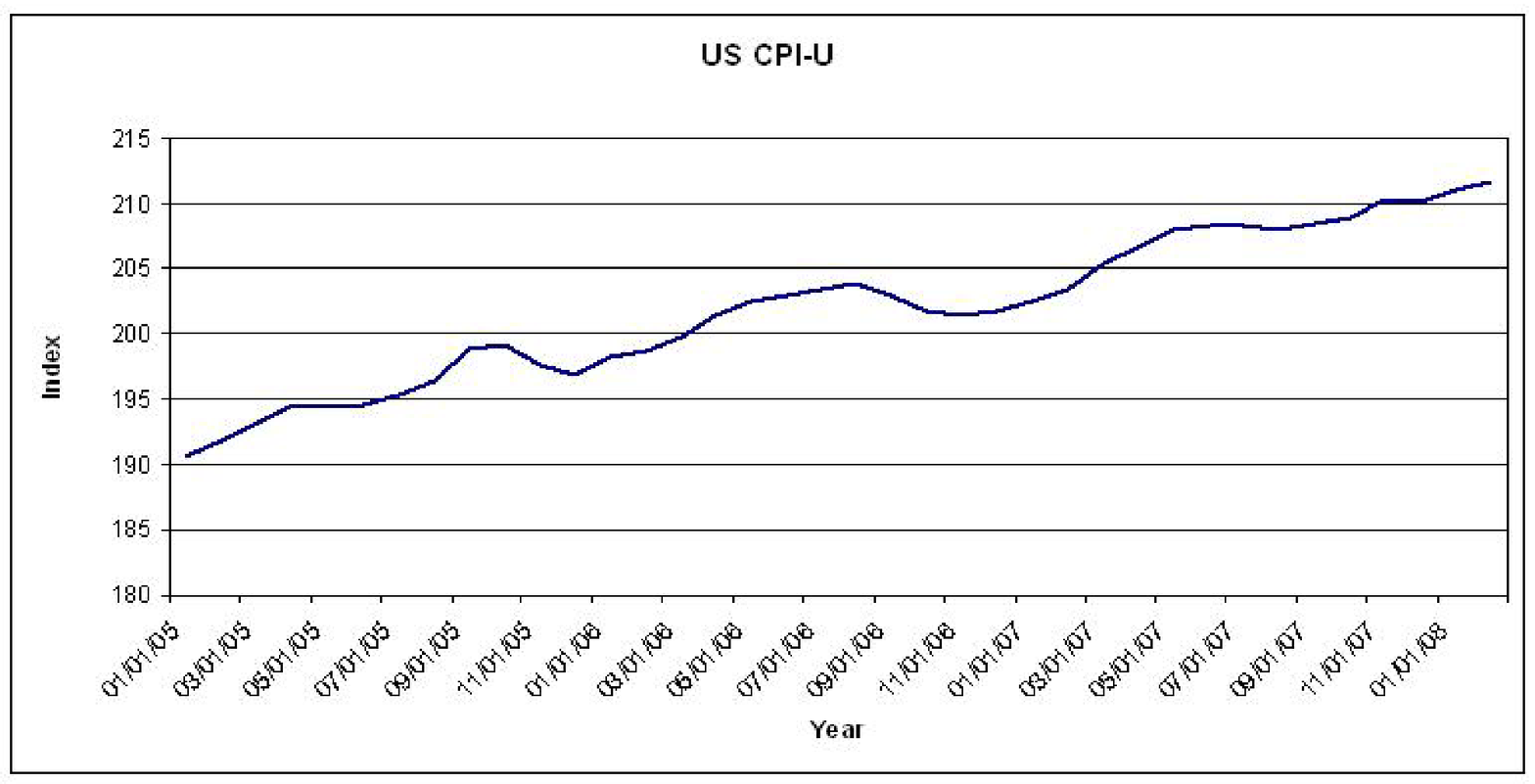}\ \ \epsfxsize=2.5in \epsfbox{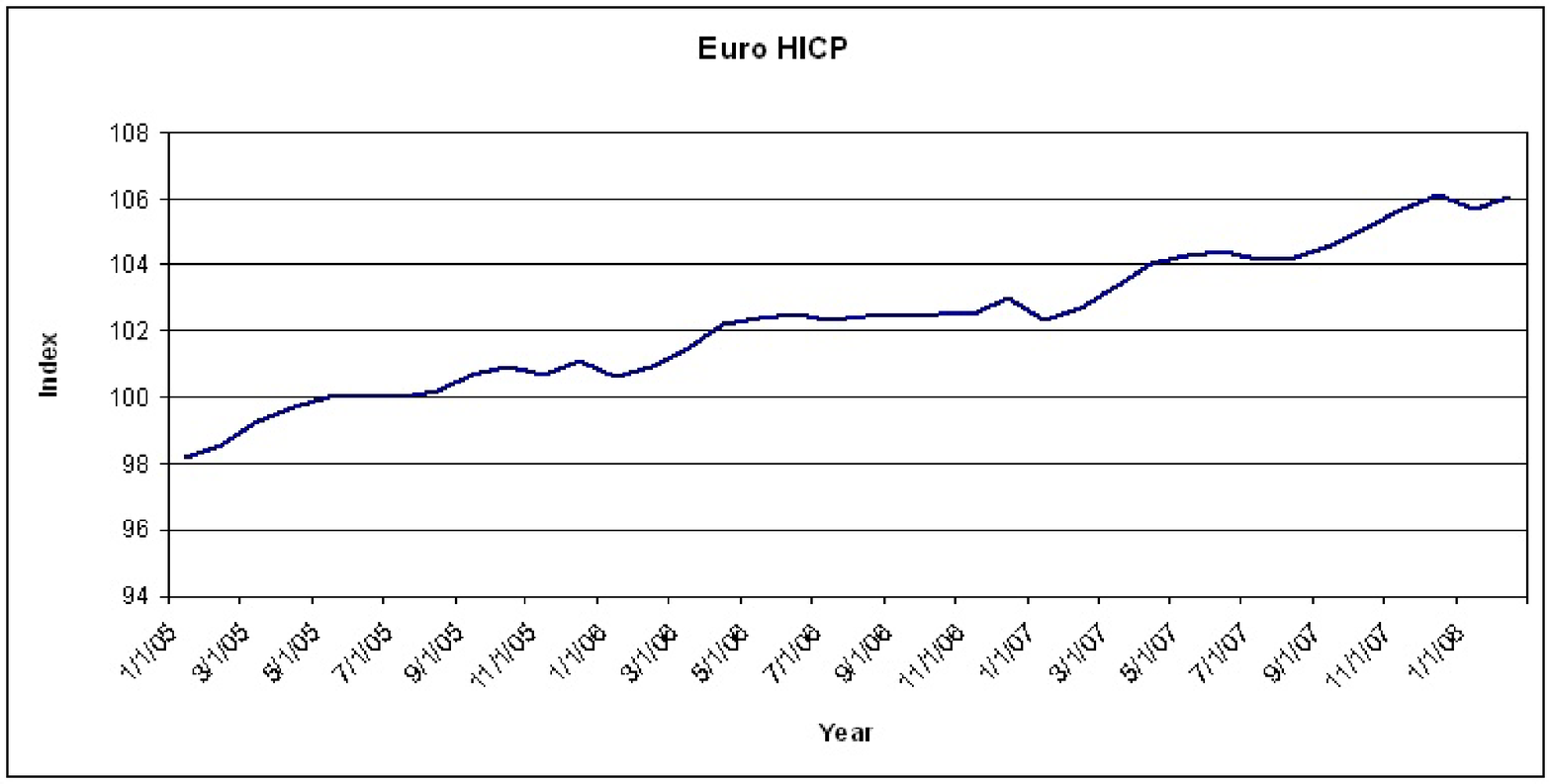}}
\centerline{{\bf Figure 1} Consumer Price Indexes of United States and Euro zone}
\vspace*{0.2in}

The inflation rate of a country is defined in terms of its CPI. Denote by $I(t)$ the CPI of time $t$, then the
inflation rate over the time period $[t, T ]$ is defined as the percentage change of the index:
\[
\hat i(t,T)={I(T)\over I(t)}-1.\]
For comparison purpose, we will more often use annualized inflation rate,
\[
i(t,T)={1\over T-t}\left({I(T)\over I(t)}-1\right).\]
Suppose the
limit of the annualized inflation rate exists for $T\rightarrow t$
from above, we obtain the so-called instantaneous inflation rate,
$i(t)$, which will be used largely for mathematical and financial arguments instead
of modeling. An important feature that distinguishes inflation rates from
interest rates is that the former can be either positive or
negative, while the latter have to be positive or otherwise we have
a situation of arbitrage.

The dollar-denominated inflation-link securities have been predominately
represented by Treasury Inflation Protected Securities (TIPS),
followed by zero-coupon inflation-indexed swap (ZCIIS) and year-on-year
inflation-indexed swap (YYIIS). In recent years, caps, floors and
swaptions on inflation rates have been gaining popularity.
The TIPS are issued by the Treasury Department of the
United States and the governments of several major industrial nations, while other derivatives are offered and traded in
the OTC markets. We emphasize here that, unlike the market model currently in use, ZCIIS are taken as the underlying securities of the inflation derivatives markets and used for the construction of ``inflation forward rates".

To understand the roles of the basic securities in model building, we need set up the economy in
mathematical terms. The uncertain
economy is modeled by a filtered probability space $(\Omega, \calF,
\{\calF_t\}_{t\in [0, \tau]}, Q)$ for some $\tau>0$, where $Q$ is the risk neutral
probability measure under the uncertain economical environment, which can be defined in a usual way in
an arbitrage-free market (Harrison and Krep, 1979; Harrison and Pliska, 1981), and
the filtration $\{\calF_t\}_{t\in [0, \tau]}$ is generated by a
$d$-dimensional $Q$ Brownian motion ${\mbox{\boldmath
$Z$}}=\{{\mbox{\boldmath $Z$}_t}: t\geq 0\}$.

Next, we will spend some length to describe these
inflation-linked securities.

\subsection{TIPS}
TIPS are coupon bonds with fixed coupon rates but floating
principals, and the latter is adjusted according to the inflation rate over
the accrual period of a coupon payment. Note that typically there is a floor
on the principal value of a TIPS, which is often the initial principal
value. The existence of floors, as a matter of fact, turns TIPS into coupon bonds
with embedded options. So the rigorous pricing of TIPS needs a
model.

Note that the CPI index is measured with a two-month lag. Yet this lagged
index plays the role of the current index for the principal adjustments of TIPS and
the payoff calculations of inflations derivatives. For pricing purpose, lagging or not makes
no difference. With this understanding in mind,
we will treat the lagged index as the current index throughout the paper.

\subsection{ZCIIS}
The zero-coupon inflation-indexed swap (ZCIIS) is a swap contract
between two parties with a single exchange of payments. Suppose that
the contract was initiated at time $t$ and will be expired at $T$,
then the payment of one party equals to a notional value times to
the inflation rate over the contract period, i.e.
\[
Not.\times \hat i(t,T),\]
while the counterparty makes a
fixed payment in the amount
\[
Not.\times \left((1+K(t,T))^{T-t}-1\right).\]
Here, $Not.$ is the notional
value of the contract and $K(t,T)$ is the quote for the contract.
Because the value of the ZCIIS is zero at initiation, ZCIIS directly
renders the price of the so-called real zero-coupon bond which pays
inflation adjusted principal:
\begin{equation}
\begin{split}
P_R(t,T)&=E^Q\left[\left . e^{-\int^T_tr_sds}{I(T)\over
I(t)}\right|\calF_t\right]=P(t,T)(1+K)^{T-t}.
\end{split}\label{e201}
\end{equation}
Here, $P(t,T)$ is the nominal discount factor
from $T$ back to $t$.
For real zero-coupon bonds with the same maturity date $T$ but
an earlier issuance date, say, $T_0<t$, the price is
\begin{equation}
\begin{split}\label{e202}
P_R(t,T_0,T)&=E^Q\left[\left . e^{-\int^T_tr_sds}{I(T)\over
I(T_0)}\right|\calF_t\right] ={I(t)\over I(T_0)}P_R(t,T).
\end{split}
\end{equation}
We emphasize here that $P_R(t,T_0,T)$, but not $P_R(t,T)$, is
treated as the time $t$ price of a traded security. The latter is
merely the initial price of a new security.

For modeling inflation-rate derivatives, we will take the term
structure of real zero-coupon bonds, $P_R(t,T_0,T)$, for a fixed $T_0\leq t$ and
for all $T\geq t$,
as model primitives. Let us explain why we use index $R$ instead
of $I$ for real zero-coupon bond defined in (\ref{e202}). This
price alone actually carries information on real interest rates instead of
inflation rates in the future. In fact, let $i(t)$ denote the
instantaneous inflation rate,
then it relates to CPI by
\begin{equation} \label{e206}
{I(T)\over I(T_0)}=e^{\int^T_{T_0}i(s)ds}.
\end{equation}
Plugging (\ref{e206}) into (\ref{e202}) yields, by Fisher's
equation (Fisher, 1930; also see Cox, Ingersoll and Ross, 1985),
\begin{equation}\label{207}
r(t)=R(t)+i(t),
\end{equation}
where $R(t)$ is the real interest rate,
we have
\begin{equation}
\begin{split}\label{e207}
P_R(t,T_0,T)&={I(t)\over
I(T_0)}E^Q\left[e^{-\int^T_t(r_s-i(s))ds}|\calF_t\right] \\
&={I(t)\over
I(T_0)}E^Q\left[e^{-\int^T_tR_sds}|\calF_t\right],
\end{split}
\end{equation}
According to (\ref{e207}), the real zero-coupon bond implies
the discount factor associated to real interest rate. This is the reason why we use the subindex ``$R$" for the price.

We emphasize here that we do not need the real interest rate for
modeling or pricing purpose, which is unobservable and
thus is not a good candidate for state variables.

\subsection{YYIIS}
Year-on-year inflation-indexed swaps are contracts to swap
an annuity against a sequence of floating payments indexed to
inflation rates over future periods. The fixed-leg payments of a
YYIIS are $Not.\D\phi_iK,i=1,2,\ldots,N_{x}$, where $\D\phi_i$ is the year fractions between two
consecutive payments, while the floating-leg
payments are of the form
\begin{equation*}
\begin{split}
& Not.\left({I(T_j)\over I(T_{j-1})}-1\right),
\end{split}
\end{equation*}
and are made at time $T_j, j=1,2,\ldots,N_{f}$. Note that the payment gaps $\D\phi_i=\phi_i-\phi_{i-1}$ and $\D T_j=T_j-T_{j-1}$ can be different, and the term for payment swaps are the same, i.e., $\sum^{N_x}_{i=1}\Delta\phi_i=\sum^{N_f}_{j=1}\Delta T_j$.
The price of the
YYIIS equals to the difference in values of the fixed and
floating legs. The former can be calculated by discounting, yet the
later involves the evaluation of an expectation:
\begin{equation*}
\begin{split}
V_{float}^{(j)}(t)&=Not.E^{Q}\left[\left. e^{-\int^{T_j}_tr_sds}\left({I(T_j)\over
I(T_{j-1})}-1\right)\right|\calF_t\right] .
\end{split}
\end{equation*}
The valuation of the floating leg will again need a model.

\subsection{Inflation Caps and Floors}
An inflation cap is like a YYIIS with optionality: with the same payment frequency, payments are made
only when a netted cash flow to the payer (of the fixed leg) is
positive, corresponding to cash flows of the following form to the
cap holder
\[
Not.\D T_i\left[{1\over \D T_i}\left({I(T_i)\over I(T_{i-1})}-1\right)-K\right]^+,
i=1,\ldots,N.\]
Accordingly, the cash flows of an inflation floor
is
\[
Not.\D T_i\left[K-{1\over \D T_i}\left({I(T_i)\over I(T_{i-1})}-1\right)\right]^+,
i=1,\ldots,N.\]
Apparently, the pricing of both caps and floors requires a
model as well.

\subsection{Inflation Swaptions}
An inflation swaption is an option to enter into a YYIIS swap in the
future. At maturity of the option, the holder of the option should
enter into the underlying YYIIS if the option is in-the-money. Up to now the pricing of the
inflation swaps has been model dependent, but the situation should change with the
establishment of the theory of this paper.

\section{The Market Model}

\subsection{Inflation Discount Bonds}
We construct models based on the dynamics of the term structures of nominal and real bonds, $\{P(t,T), \forall T\geq t\}$ and
$\{P_R(t,T_0,T),\forall T\geq t\geq T_0\}$, two sequences of tradable securities.
Under the risk neutral measure $Q$, $P(t,T)$ is assumed to follow the
lognormal process
\begin{equation} \label{e310}
dP(t,T) = P(t,T)\left(r_tdt + \Sigma(t,T)\cdot d{\mbox{\boldmath
$Z$}}_t\right),
\end{equation}
where $r_t$ is the risk-free nominal (stochastic) interest rate, $\Sigma(t,T)$ is a
$d$-dimensional volatility vector of $P(t,T)$ and ``$\cdot$"
means scalar product. We shall assume that $\Sigma(t,T)$ is a
sufficiently regular deterministic function on $t$ so that the SDE
(\ref{e310}) admits a unique strong solution. Note that $\Sigma(t,T)$
can be an ${\cal F}_t$-adaptive (stochastic) function. Furthermore, we also
assume $\Sigma_T(t,T)=\frac{\partial \Sigma(t,T)}{\partial T}$ exists
and
$E^Q[\int_0^T{\|\Sigma_T(s,T)\|^2ds}] < \infty$.

By using Ito's lemma, we have the following process for $\ln P(t,T)$:
\begin{equation}\label{e311}
d \ln P(t,T) = \left(r_t - \frac{\|\Sigma(t,T)\|^2}{2}\right)dt +
\Sigma(t,T)\cdot d{\mbox{\boldmath $Z$}}_t,
\end{equation}
where $\|x\|^2 = x\cdot x$ for $x \in \Bbb R^d$. Differentiating
equation (\ref{e311}) with respect to the maturity $T$, we have
\begin{equation}\label{HJM3}
df(t,T) = \Sigma_T(t,T)\cdot \Sigma(t,T)dt - \Sigma_T(t,T)\cdot
d{\mbox{\boldmath $Z$}}_t,
\end{equation}
where $f(t,T)= -\frac{\partial \ln P(t,T)}{\partial T}$ is the
nominal instantaneous forward rate of maturity $T$. Equation
(\ref{HJM3}) is the well-known Heath-Jarrow-Morton equation (Heath
\etal 1992) for term structure of nominal interest rates, which states that, under the risk neutral measure $Q$,
the drift term of the forward rate is a function of its
volatility.

The dynamics of $P_R(t,T_0,T)$ under the risk neutral
measure $Q$, meanwhile, is also assumed to be lognormal:
\begin{equation}\label{e375}
dP_R(t, T_0 ,T) = P_R(t, T_0 ,T)\left(r_tdt + \Sigma_R(t,T)\cdot
d{\mbox{\boldmath $Z$}}_t\right),
\end{equation}
where $\Sigma_R(t,T)$ is the $d$-dimensional volatility vector of
$P_R(t, T_0 ,T)$ and satisfies the similar regularity conditions as
$\Sigma(t,T)$ does. One can easily justifies that, using (\ref{e201}) and
(\ref{e202}), $\Sigma_R(t,T)$ should be independent of $T_0$.

To define the term structure of inflation rates, we first introduce the
notion of discount bond or discount factor associated to inflation rate, using $P(t,T)$ and $P_R(t,T)$, the nominal and real discount bond prices or factors.

{\bf Definition 1}: The {\it discount bond associated to inflation rate\/} is defined by
\begin{equation}
P_I(t,T)\defn {P(t,T)\over P_R(t,T)}. \label{e209}
\end{equation}
Here, ``$\defn$" means ``being defined by".

Alternatively, with $P_I(t,T)$ and $P_R(t,T)$, we effectively factorize the nominal
discount factor into real and inflation discount factors:
\begin{equation}
P(t,T)= { P_R(t,T)P_I(t,T)}.
\end{equation}
Note that neither $P_I(t,T)$ nor $P_R(t,T)$ is a price of a tradable
security\footnote{$P_R(t,T)$ is treated as the price of a zero-coupon
bond of a virtue ``foreign currency" by Jarrow and Yildirim
(2003).}, but they both are observable.
For later uses, we denote
\begin{equation}
\begin{split}
P_I(t,T_0,T)={P(t,T)\over P_R(t,T_0,T)},
\end{split}\label{e313}
\end{equation}
so there is
\begin{equation}
\begin{split}
P_I(t,T)&={I(t)\over I(T_0)}{P_I(t,T_0,T)}.
\end{split}
\label{e210}
\end{equation}
Note that $P_I(t,T_0,T)$ as well as $P_I(t,T)$ are defined for $t>T$ as well, through a
constant extrapolation:
\begin{equation}
\begin{split}
P_I(t,T_0,T)&=P_I(T,T_0,T), \quad \forall t\geq T.
\end{split}
\label{e211}
\end{equation}
This is because that $P_I(t,T_0,T)$ is the ratio between $P(t,T)$ and $P_R(t,T_0,T)$. At time $T$, both securities mature into money market account, and the ratio stays unchange since then.

\subsection{Market Model for Inflation Derivatives}
It can be seen that the cash flows of several major
inflation-indexed instruments, including YYIIS, inflation caplets and floorlets, are
expressed in terms of forward inflation term rates (or simple inflation rates).
We define a inflation forward rate
as the {\it return implied by the inflation discount factor\/}.

{\bf Definition 2}: The {\it inflation forward rate\/} for a future period $[T_1,T_2]$
seen at time $t\leq T_2$ is defined by
\begin{equation}
\begin{split}
f^{(I)}(t,T_1,T_2)&={1\over (T_2-T_1)}\left({P_I(t,T_1)\over P_I(t,T_2)}-1\right).
\end{split}\label{e527}
\end{equation}

It can be seen easily that 1) the definition for the inflation forward rates is equivalent to
\begin{equation}
\begin{split}
f^{(I)}(t,T_1,T_2)
&={1\over (T_2-T_1)}\left({P_I(t,T_0,T_1)\over
P_I(t,T_0,T_2)}-1\right),
\end{split}\label{e528}
\end{equation}
and 2) at $T_2$, the fixing date, we will have the convergence of the inflation forward rate to
the spot inflation rate:
\begin{equation}
\begin{split}
f^{(I)}(T_2,T_1,T_2)={1\over T_2-T_1}\left({I(T_2)\over
I(T_1)}-1\right).
\end{split}\label{FI5}
\end{equation}
As a result, the payoff functions of several major derivatives can now be written in terms of inflation forward rates. Derivatives pricing can be made convenient provided we have a simple and analytical tractable model for the inflation forward rates.

We make a remark that, through straightforward derivations, one will see that the definition of inflation forward rates by (\ref{e527}) is actually the same as one of the definitions,
$Y_i(t)$, in Mercurio and Moreni (2009).
We emphasize here that the inflation forward rate so defined is the unique fair rate seen at the time $t$ for a $T_1$-expiry forward contract on the inflation rate over the future period $[T_1,T_2]$. The justification of the next proposition is given in the appendix.

{\bf Proposition 1}: The time-$t$ forward price to purchase a real bond with maturity $T_2$ at time $T_1$ such that $t\leq T_1\leq T_2$ is
\begin{equation}
\begin{split}
F_R(t,T_1,T_2)\defn{P_R(t,T_0,T_2)\over P_R(t,T_0,T_1)}.
\end{split}\label{e531}
\end{equation}

Based on the above proposition, we can show that the inflation forward rate defined in (\ref{e527}) is the only arbitrage-free rate for forward contracts. Let $f$ be the no-arbitrage strike rate for a $T_2$-expiry forward contract on the inflation rate over $[T_1,T_2]$ that pays $(T_2-T_1)(f^{(I)}(T_2,T_1,T_2)-f)$. We will do the following sequence of transactions.
\begin{enumerate}
\item At time $t$,
\begin{enumerate}
\item Short the $T_1$-expiry forward contract on $f^{(I)}(T_2,T_1,T_2)$;
\item Long a $T_1$-expiry forward contract with strike price $F_R(t,T_1,T_2)$ on one unit of the real bond with tenor $T_2> T_1$;
\item Short the $T_2$-maturity Treasury discount bond and long the $T_1$-maturity Treasury discount bond with an equal dollar value of $F_R(t,T_1,T_2)P(t,T_1)$.
\end{enumerate}
\item At time $T_1$, exercise the $T_1$-expiry forward contract by purchasing the real bond for $F_R(t,T_1,T_2)$ dollars, the proceed from the $T_1$-maturity Treasury discount bond.
\item  At time $T_2$, close out all positions.
\end{enumerate}
At $T_2$, we end up with the following net value of the sequence of zero-net transactions:
\begin{equation*}
\begin{split}
P\&L=&(T_2-T_1)[f-f^{(I)}(T_2,T_1,T_2)]
+{I(T_2)\over I(T_1)}-{F_R(t,T_1,T_2)P(t,T_1)\over P(t,T_2)} \\
=&(T_2-T_1)[f-f^{(I)}(T_2,T_1,T_2)]
+{I(T_2)\over I(T_1)}-{P_I(t,T_0,T_1)\over P_I(t,T_0,T_2)} \\
=&(T_2-T_1)[f-f^{(I)}(t,T_1,T_2)].
\end{split}
\end{equation*}
Apparently, arbitrage occurs if $f\neq f^{(I)}(t,T_1,T_2)$.

Being a $T_1$-forward price of a tradable security, $F(t,T_1,T_2)$ should be a lognormal martingale under the $T_1$-forward measure whose volatility is the difference of those of $P_R(t,T_0,T_2)$ and $P_R(t,T_0,T_1)$, i.e.,
\begin{equation}
\begin{split}
{dF_R(t,T_1,T_2)\over F_R(t,T_1,T_2)}
&=\left(\Sg_R(t,T_2)-\Sg_R(t,T_1)\right)^T(d{\bf Z}_t-\Sg(t,T_1)dt).
\end{split}\label{e532}
\end{equation}
Note that, in general, $d{\mbox{\boldmath
$Z$}}_t-\Sg(t,T)dt$  is (the differential of) a Brownian
motion under the so-called $T$-forward measure, $Q_T$, which is defined by the Radon-Nikodym derivative
\[
\left. {d Q_T\over d Q}\right|_{{\cal F}_t}={P(t,T)\over B(t)P(0,T)},\]
where $B(t)=exp(\int^t_0r_sds)$ is the unit price of money market account.

There is an important implication by (\ref{e532}). Based on the risk neutral dynamics of $P_R(t,T_0,T)$, there is also
\begin{equation}
\begin{split}\label{e5341}
{dF_R(t,T_1,T_2)\over F_R(t,T_1,T_2)}=\left(\Sg_R(t,T_2)-\Sg_R(t,T_1)\right)^T(d{\bf Z}_t-\Sg_R(t,T_1)dt).
\end{split}
\end{equation}
The coexistence of equations (\ref{e532}) and (\ref{e5341}) poses a constraint on the volatility functions on the real bonds.

{\bf Proposition 2 (Consistency condition)}: For arbitrage pricing, the volatility functions of the real bonds must satisfy the following condition:
\begin{equation}
\begin{split}
\left(\Sg_R(t,T_2)-\Sg_R(t,T_1)\right)\cdot (\Sg(t,T_1)-\Sg_R(t,T_1))=0.
\end{split}\label{e318}
\end{equation}

Literally, the consistency condition is equivalent to say that
\[
Cov \left(d\left({P_R(t,T_0,T_2)\over P_R(t,T_0,T_1)}\right),d\left({P(t,T_1)\over P_R(t,T_0,T_1)}\right)\right)=0.\]
While an intuitive interpretation is not available at this point, we can at least show that the consistency condition holds provided that
the real forward rate and the inflation rate are uncorrelated, because then there will be
\[
{P(t,T_1)\over P_R(t,T_0,T_1)}=E^Q_t\left[e^{-\int^{T_1}_ti_sds} \right],\]
while
\[
f_R(t,T_1,T_2)={1\over \D T}\left({P_R(t,T_0,T_1)\over P_R(t,T_0,T_2)}-1\right) \]
is the real forward rate.
Let
\[\Sg_I(t,T)=\Sg(t,T)-\Sg_R(t,T)\]
denote the volatility of $P_I(t,T_0,T)$.
Divide (\ref{e318}) by $(T_2-T_1)$ and let $T_2\rightarrow T_1=T$, we then end up with
\begin{equation}
\begin{split}
\dot\Sg_R(t,T)\cdot \Sg_I(t,T)=0.
\end{split}\label{e319}
\end{equation}
This version of consistency of consistency condition will be use later to derive a Heath-Jarrow-Morton type model with instantaneous inflation rates.

Using (\ref{e310}) and (\ref{e532}), we can derive the dynamics of the inflation forward rate $f^{(I)}(t,T_1,T_2)$. For
generality, we let $T=T_2$, $\D T=T_2-T_1$, we then can cast
(\ref{e527}) into
\[
f^{(I)}(t,T-\D T,T)+{1\over \D T}={1\over \D T}{F_R(t,T-\D T,T)P(t,T-\D T) \over
P(t,T)}.\]
The dynamics of $f^{(I)}(t,T-\D T,T)$ follows from those of
$F_R$ and $P$'s (and thus is left to readers).

{\bf Proposition 3}. Under the risk neutral measure, the governing equation for the simple inflation forward rate
is
\begin{equation}
\begin{split}
&d\left(f^{(I)}(t,T-\D T,T)+{1\over\D T}\right)\\
=&\left(f^{(I)}(t,T-\D T,T)+{1\over\D T}\right) \left\{\g^{(I)}(t,T)\cdot
(d{\mbox{\boldmath
$Z$}}_t-\Sg(t,T)dt)\right\} ,
\end{split}
\label{e533}
\end{equation}
where
\begin{equation*}
\begin{split}
\g^{(I)}(t,T)&=\Sg_I(t,T-\D T)-\Sg_I(t,T)
\end{split}
\end{equation*}
is the percentage volatility of the displaced inflation forward rate.

The displaced diffusion dynamics (\ref{e533}) for the simple inflation rates has at least two desirable
features. First, it allows the inflation rates to take both positive and negative values,
reflecting the economical environment of either inflation or deflation. There is a lower bound, $-1/\D T$, on the inflation rate, which effectively prevents the prices of goods from becoming negative. Second, it is analytical
tractable for derivatives pricing. For the purpose of derivatives pricing, we will use (\ref{e533})
in conjunction with a term structure model for nominal interest rates, preferably a model with simple compounding nominal forward rates. As such, the choice for
a term structure model with simple nominal forward rates points to the LIBOR market model (Brace et. al, 1997; Jamshidian, 1997; Miltersen and Sandmann, 1997), which is the benchmark model for nominal interest rates and has has a number of desirable features for a term structure model.

We are now ready to propose a comprehensive market model for inflation rates. The
state variables consist of two streams of spanning forward rates and
inflation forward rates, $f_j(t)\defn f(t,T_j,T_{j+1})$ and
$f^{(I)}_j(t)\defn f^{(I)}(t,T_{j-1},T_j), j=1,2,\ldots,N$, that follow the
following dynamics:
\begin{equation}
\left\{
\begin{split}\label{e534}
df_j(t)&=f_j(t)\g_j(t)\cdot \left(d{\bf Z}_t-\Sg_{j+1}(t)dt\right), \\
d\left(f^{(I)}_j(t)+{1\over\D T_j}\right)&=\left(f^{(I)}_j(t)+{1\over\D T_j}\right) \g^{(I)}_j(t)\cdot (d{\bf Z}_t-\Sg_j(t)dt),
\end{split}\right .
\end{equation}
where
\begin{equation*}
\begin{split}
\Sg_{j+1}(t)&=-\sum_{k=\eta_t}^{j}{\D T_{k+1} f_k(t) \over 1+\D T_{k+1} f_k(t)}\g_{k}(t),
\end{split}
\end{equation*}
and
\begin{equation*}
\begin{split}
& \eta_t=\min\{i|T_i>t\}.
\end{split}
\end{equation*}

As we shall see shortly, with the
lognormal processes for nominal and inflation forward rates, the
pricing of major inflation derivatives can be made very
convenient.

The market model just developed lends itself for further extensions. In its current form,
the model cannot accommodate implied volatility smiles or skews. For these ends, we may
incorporate additional risk factors like jumps and/or stochastic volatilities into the
equations. In section 6, we will make a brief discussion on possible extensions of the market model.

\subsection{The Extended Heath-Jarrow-Morton Model}
Analogously to the introduction to nominal forward rates, we now
introduce the instantaneous inflation forward rates, $f^{(I)}(t,T)$, through
\begin{equation}
\begin{split}
f^{(I)}(t,T)&=-\frac{\partial \ln P_I(t,T)}{\partial T}, \quad \forall T\geq t,
\end{split}\label{e314}
\end{equation}
or
\[
P_I(t,T)=e^{-\int^T_tf^{(I)}(t,s)ds}.\]
According to (\ref{e313}), we can express the instantaneous forward rate as
\begin{equation*}
\begin{split}
f^{(I)}(t,T)
&=-\frac{\partial \ln P_I(t,T_0,T)}{\partial T}={\p \ln \left({P_R(t,T_0,T) \over P(t,T)}\right)\over {\p T}},\quad \forall T\geq t.
\end{split}
\end{equation*}
The dynamics of
$f^{(I)}(t,T)$, therefore, follows from those of $P(t,T)$ and $P_R(t,T_0,T)$.
By the Ito's lemma, we have
\begin{equation}
\begin{split}
-d\ln P_I(t,T_0,T)&= d\ln\left(P_R(t,T_0,T)\over P(t,T)\right)\\
&=-{1\over 2}\|\Sg_I(t,T)\|^2dt-\Sg_I^T(t,T)\left(d{\bf W}_t-\Sg(t,T)dt\right).
\end{split}\label{e320}
\end{equation}
Differentiating the above equation with respect to $T$ and making use of the consistency condition (\ref{e319}),
we then have
\begin{equation}\label{e321}
\begin{split}
df^{(I)}(t,T)&=-\dot\Sigma_I\cdot\left(
d{\bf Z}_t-\Sg(t,T)dt\right),
\end{split}
\end{equation}
where the overhead dots mean partial derivatives with
respect to $T$, the maturity. Equation (\ref{e321}) shows that $f^{(I)}(t,T)$ is a martingale and its dynamics is fully specified by the volatilities of the nominal and inflation forward rates. The joint equations of (\ref{HJM3}) and (\ref{e321}) constitute the
so-called extended Heath-Jarrow-Morton framework (or model) for nominal interest rates and inflation rates.

For applications of the model, we will instead first prescribe the
volatilities of forward rates and inflation forward rates, defined by
\begin{equation*}\label{e315}
\begin{split}
& \sigma(t,T) = -\dot\Sigma(t,T),\\
& \sigma^{(I)}(t,T) =-\dot\Sigma_I(t,T).
\end{split}
\end{equation*}
In terms of $\sigma(t,T)$ and $\sigma^{(I)}(t,T)$, we can expresses the
volatilities of nominal zero-coupon bonds as
\begin{eqnarray}\label{HJM10}
&& \Sigma(t,T)=-\int_t^T {\sigma(t,s)ds}, \nonumber
\end{eqnarray}
and then cast our extended HJM model in
joint equations with the forward rates and inflation forward rates:
\begin{equation}\label{HJM11}
\left\{
\begin{split}
df(t,T) &= \sigma(t,T) \cdot d{\bf Z}_t+\sigma(t,T)\cdot \left(\int_t^T {\sigma(t,s)ds}\right) dt,  \\
df^{(I)}(t,T) &= \sigma^{(I)}(t,T) \cdot d{\bf Z}_t
+\s^{(I)}(t,T)\cdot \left(\int^T_t\s(t,s) ds\right)dt.
\end{split} \right .
\end{equation}
The initial term structures of forward rates and inflation forward rates serve as inputs to the these equations.

Let us establish the connection between our model and that of Jarrow and Yildirim (2003) based on ``foreign currency analogy". The instantaneous real forward rate satisfies
\[
f_R(t,T)=f(t,T)-f^{(I)}(t,T).\]
Let
\[
\sigma_R(t,T)=-\dot\Sg_R(t,T)=\sigma(t,T)-\sigma^{(I)}(t,T).\]
Then
\[
\Sg_R(t,T)=-\int^T_t\s_R(t,s)ds+\s_I(t),\]
where $\s_I(t)$ is the volatility of the CPI index $I(t)$. Subtracting the two equations of (\ref{HJM11}) and applying the consistency condition, (\ref{e319}), we will arrive at
\begin{equation}
\begin{split}
df_R(t,T) &= \sigma_R(t,T) \cdot d{\bf Z}_t+\sigma_R(t,T)\cdot \left(\int_t^T {\sigma_R(t,s)ds}-\s_I(t)\right) dt,
\end{split} \label{e328}
\end{equation}
which is identical to the dynamics the real forward rate established by
Jarrow and Yildirim (2003) (page 342, equation (12))! Hence, our model is consistent with the model of Jarrow and Yildirim, established using ``foreign currency analogy", a very different approach. With the above results, we claim that our model and the model of Jarrow and Yildirim are two variants of the same model for inflation-rate derivatives.

We are, however, reluctant to accept ``foreign currency analogy" for the reason that we actually have $\s_I(t)=0$. The dynamics of the CPI index follows from the definition of of the CPI index, (\ref{e206}), and the Fisher's equation:
\begin{equation}\label{e0330}
dI(t)=i(t)I(t)dt=(r_t-R_t)I(t)dt,
\end{equation}
and this simple fact has long been overlooked in the literature on inflation-rate modeling. The implication is that CPI index cannot be treated as an exchange rate between the nominal and real (or virtue) economies, unless it is completely determined by the interest rates of the two economies as in (\ref{e0330}).

\subsection{Pricing of YYIIS}
The price of a YYIIS is the difference in value of the fixed leg and floating leg.
While the
fixed leg is priced as an annuity,
the floating leg is priced by discounting the expectation of each piece of
payment as
\begin{equation*}
\begin{split}
V_{float}^{(j)}(t)&=Not.P(t,T_j)E^{Q_{T_j}}_t\left[\left({I(T_j)\over I(T_{j-1})}-1\right)\right] \\
&=Not.\D T_jP(t,T_j)E^{Q_{T_j}}_t\left[f^{(I)}_j(T_j)\right] \\
&=Not.\D T_jP(t,T_j)f^{(I)}_j(t),
\end{split}
\end{equation*}
followed by a summation:
\[
V_{float}(t)=Not.\sum^{n_f}_{j=1}\D T_jP(t,T_j)f^{(I)}_j(t).\]

We result we have here differs greatly from the current practice of the market, where the pricing of YYIIS makes no use of the inflation forward rates implied by ZCIIS. In existing literatures, the pricing of YYIIS based on ZCIIS goes through a procedure of ``convexity adjustment", which is model dependent. With our result, we realize that YYIIS can and should be priced consistently with XCIIS, otherwise arbitrage opportunities will occur.

\subsection{Pricing of Inflation Caplets}
In view of the displaced diffusion processes for simple inflation forward rates, we can price a
caplet with \$1 notional value straightforwardly as follows:
\begin{equation}\label{MM5}
\begin{split}
&\D T_jE^{Q}_t\left[e^{-\int^{T_j}_tr_sds}(f^{(I)}_j(T_j)-K)^+  \right]\\
=&\D T_jP(t,T_j)E^{Q_{T_j}}_t\left[\left(\left(f^{(I)}_j(T_j)+{1\over \D {T_j}}\right)-\left(K+{1\over \D {T_j}}\right)\right)^+  \right]\\
=&\D {T_j}P(t,{T_j})\{\mu_j\Phi(d_1^{(j)}(t))-\td K_j
\Phi(d_2^{(j)}(t))\},
\end{split}
\end{equation}
where $\Phi(\cdot)$ is the standard normal accumulative distribution function, and 
\begin{equation*}
\begin{split}
&\mu_j=f^{(I)}_j+1/\D T_j, \quad \td K_j=K+1/\D T_j, \\
&d_1^{(j)}(t)={\ln{\mu_j/\td K_j}+{1\over 2}\s^2_j(t)(T_j-t)\over \s_j(t)\sqrt{T_j-t}}, \\
&d_2^{(j)}(t)=d_1^{(j)}(t)-\s_j(t)\sqrt{T_j-t},
\end{split}
\end{equation*}
with $\s_j$ to be the volatility of $\ln(f^{(I)}_j(t)+{1\over \D
{T_j}})$:
\begin{equation}
\begin{split}
\s^2_j(t)=& {1\over T_j-t}\int^{T_{j}}_t\|\g^{(I)}_j(s)\|^2ds.
\end{split}\label{e537}
\end{equation}

The inflation-indexed cap with maturity $T_N$ and strike $K$ is the
sum of a series of inflation-indexed caplets with the cash flows at
$T_j$ for $j=1,\cdots,N$. We denote by $\mbox{IICap}(t;N,K)$
the price of the inflation-indexed cap at time $t$, where $T_0 < t
\leq T_1$, with cash flow dates $T_j, j=1,\ldots,N$, and strike $K$.
Based on (\ref{MM5}), we have
\begin{equation}\label{MM6}
\begin{split}
&\mbox{IICap}(t;N,K) \\
=& \sum_{j=1}^N\Delta T_j P(t,T_j)\{\mu_j\Phi(d_1^{(j)}(t))-\td
K_j\Phi(d_2^{(j)}(t))\}.
\end{split}
\end{equation}

Given inflation caps of various maturities, we can consecutively bootstrap
$\s_j(t)$, the ``implied caplet volatilities", in either a parametric or
a non-parametric way.
With additional information on correlations between inflation rates of
various maturities, we can determine $\g^{(I)}_j$, the volatility
of inflation rates and thus fully specify the displace-diffusion dynamics for inflation forward rates.
We may also include inflation swaption prices to the input set
to specify $\g^{(I)}_j$'s.

\section{Pricing of Inflation-Indexed Swaptions}
The Year-on-Year Inflation-Indexed Swaption (YYIISO) is an option to
enter into a YYIIS at the option's maturity. Base on our market model (\ref{e534}),
we will show that
a forward inflation swap rate with a displacement is a martingale under
a usual nominal forward swap measure. Instead of
assuming lognormality for the inflation swap rate as in
Hinnerich (2008), we justify that the displaced inflation swap rate
is a Gaussian martingale and for which we produce a lognormal
dynamics by ``freezing coefficients''. The closed-form pricing of
the swaptions then follows.

Next, let us derive the expression for inflation swap rate. Without loss of generality,
we assume the same cash flow frequency for both fixed
and floating legs. The value of a payer's YYIIS over the period
$[T_m,T_n]$ at time $t\leq T_m$ for a swap rate $K$ is
given by
\begin{equation}\label{IIS1}
\begin{split}
Y_{m,n}(t,K)
&= \sum_{i=m+1}^n{ \Delta T_i
P(t,T_i)E^{Q_{T_i}}_t\left[\frac{1}{\Delta T_i}\left(\frac{I(T_i)}
{I(T_{i-1})}-1\right)-K\right]}  \\
&= \sum_{i=m+1}^n{ \Delta T_i
P(t,T_i)E^{Q_{T_i}}_t\left[f^{(I)}_i(T_i)-K\right]} \\
&= \sum_{i=m+1}^n{ \Delta T_i
P(t,T_i)\left[f^{(I)}_i(t)-K\right]}.
\end{split}
\end{equation}
The forward swap rate at $t$, denoted by $S_{m,n}(t)$, is defined as the value of
$K$ which makes the value of the swap, $Y_{m,n}(t,K)$, equal to 0.
So,
\begin{eqnarray}\label{IIS2}
S_{m,n}(t) &=& \frac{\sum_{i=m+1}^n{\Delta T_i
P(t,T_i)f^{(I)}_i(t)}}{\sum_{i=m+1}^n{\Delta
T_i P(t,T_i)}},
\end{eqnarray}
or, more preferably,
\begin{equation}\label{e640}
\begin{split}
 & S_{m,n}(t)+ \frac{1}{\Delta T_{m,n}} \\
=& \frac{\sum_{i=m+1}^n{\Delta T_i
P(t,T_i)\left[f^{(I)}_i(t)+\frac{1}{\Delta
T_i}\right]}}{\sum_{i=m+1}^n {\Delta T_i P(t,T_i)}} \\
=& \sum_{i=m+1}^n {\o_i(t)\mu_i(t)},
\end{split}
\end{equation}
where
\[
\o_i(t)={\D T_iP(t,T_i)\over A_{m,n}(t)} \quad \mbox{ and }\quad
A_{m,n}(t)=\sum_{i=m+1}^n\Delta T_i P(t,T_i),\]
and
\[{1\over \D T_{m,n}}=\sum^n_{i=m+1}\o_i(t){1\over \D T_i}.\]
We have the following results on the dynamics of the swap rate.

\textbf{Proposition 5.} The displaced forward swap rate
$S_{m,n}(t)+ \frac{1}{\Delta T_{m,n}}$ is a martingale under the
measure $Q_{m,n}$ corresponding to the numeraire $A_{m,n}(t)$.
Moreover,
\begin{equation}\label{e642}
\begin{split}
& d\left(S_{m,n}(t)+\frac{1}{\Delta T_{m,n}}\right)=\left(S_{m,n}(t)+\frac{1}{\Delta T_{m,n}}\right) \\
&\times \sum_{i=m+1}^n \left[\a_i(t)
\g^{(I)}_i(t)+(\a_i(t)-w_i(t))\Sigma_i(t)\right]\cdot\dbz^{(m,n)}_t,
\end{split}
\end{equation}
where $\dbz^{(m,n)}_t$ is a $Q_{m,n}$-Brownian motion, and
\begin{equation*}
\begin{split}
\a_i(t)={\o_i(t)\mu_i(t)\over \sum_{j=m+1}^n \o_j(t)\mu_j(t)}. \quad \Box
\end{split}
\end{equation*}

The martingale property is easy to see because it is the relative value between its
floating leg and an annuity, both are tradable. The proof of (\ref{e642}) is
supplemented in the appendix.

By freezing coefficients of
appropriately, we can turn (\ref{e642}) into a lognormal process. We proceed as
follows. Conditional on ${\cal F}_t$, we cast (\ref{e642}) for $s\geq t$ into
\begin{equation}\label{IIS15}
d\left(S_{m,n}(s)+\frac{1}{\Delta T_{m,n}}\right) =
\left(S_{m,n}(s)+\frac{1}{\Delta T_{m,n}}\right)\gamma_{m,n}^{(I)}(s)\cdot\dbz^{(m,n)}_s,
\end{equation}
where
\begin{equation*}
\begin{split}
\gamma_{m,n}^{(I)}(s) &= \sum_{i=m+1}^n \left[\a_i(t)
\g^{(I)}_i(s)+(\a_i(t)-w_i(t))\Sigma_i(s)\right], \\
\Sigma_j(s)&=-\sum^j_{k=\eta_t}{\D T_{k+1}f_k(t)\over 1+\D
T_{k+1}f_k(t)}\g_k(s).
\end{split}
\end{equation*}
As a result of freezing coefficients selectively, the volatility function $\gamma_{m,n}^{(I)}(s)$ is now deterministic, which paves the way for closed-form pricing of swaptions.

Now we are ready to price swaptions.
Consider a $T_m$-expiry
YYIISO with underlying YYIIS over the period $[T_m,T_n]$ and strike
$K$, its value, denoted the price by YYIISO$(t, T_m, T_n, K)$ at time $t \leq
T_m$, then,
\begin{equation}\label{IIS16}
\begin{split}
& \mbox{YYIISO}(t,T_m,T_n) \\
=&E^Q_t[e^{-\int_t^{T_m}r_sds}A_{m,n}(T_m)(S_{m,n}(T_m)-K)^{+}  ]
 \\
=& A_{m,n}(t)E^{Q_{m,n}}_t[(S_{m,n}(T_m)-K)^{+}]  \\
=& A_{m,n}(t)E^{Q_{m,n}}_t\left[\left[\left(S_{m,n}(T_m)+\frac{1}{\Delta T_{m,n}}\right)-\left(K+\frac{1}{\Delta T_{m,n}}\right)\right]^{+} \right]  \\
=& A_{m,n}(t) \left[\left(S_{m,n}(t)+\frac{1}{\Delta
T_{m,n}}\right)\Phi(d_1^{(m,n)})-\td K_{m,n} \Phi(d_2^{(m,n)})\right],
\end{split}
\end{equation}
where
\begin{equation*}
\begin{split}
\td K_{m,n} =& K + \frac{1}{\Delta T_{m,n}}, \\
d_1^{(m,n)} =& \frac{\ln\left(S_{m,n}(t)+1/\Delta
T_{m,n}\right)/\td K_{m,n} +{1\over 2}
\sigma_{m,n}^2(t)(T_m-t)}{\sigma_{m,n}(t)\sqrt{T_m-t}}, \\
d_2^{(m,n)} =& d_1^{(m,n)} - \sigma_{m,n}(t)\sqrt{T_m-t}, \\
\sigma_{m,n}(t) =&{1\over T_m-t} \int^{T_m}_t\|\gamma^{(I)}_{m,n}(s)\|^2ds.
\end{split}
\end{equation*}
In (\ref{IIS16}), we freeze $\o_i(s)$ at $s=t$ for evaluating $\frac{1}{\Delta
T_{m,n}}$. Because $\a_j$'s are in terms of $\mu_j(t)$'s, we must
have already obtained $\mu_j(t)$'s before applying the pricing formula.

Treatments of freezing coefficients similar to what we did to (\ref{e642}) are popular in the industry, and they are often very
accurate in many applications. A rigorous analysis on the error estimation of such approximations, however, is still pending. For some insights about the magnitude of errors, we refer to Brigo et al.
(2004).

Finally in this section we emphasize that the price formula (\ref{IIS16}) implies a hedging strategy for the swaption. At ant time $t$,
the hedger should long $\Phi(d_1^{(m,n)})$ units of the underlying inflation swap for hedging. Proceeds from buying or selling the swap may go in or go out of a money market account.

\section{Calibration of the Market Model}
A comprehensive calibration of the inflation-rate model (\ref{e534}) means simultaneous determination of volatility vectors for nominal and inflation forward rates, based on inputs of term structures and prices of benchmark derivatives. This task, luckily, can be achieved by divide-and-conquer: the LIBOR model for nominal interest rates can be calibrated in advance using only the LIBOR data, then the market model for inflation rates can be calibrated separately in a similar way,
making use of the data of inflation derivatives.

Before calibration, we need to build the spot term structure of inflation rates, using (\ref{e527}).
For a comprehensive calibration of the market model for inflation rates, we may need to match the market prices
of a set 
inflation caps/floors and inflation-rate swpations. That is, the input set consists of
\[
\{\s_j\}  \quad\mbox{and}\quad \{\s_{m,n}\}.\]
In addition, we may need to input the correlations amongst inflations rates and
between inflation rates and interest rates.
Mathematically, a comprehensive calibration amounts to solving the following joint equations
\begin{equation}
\begin{split}
&\s_j^2(T_j-t)=\int^{T_j}_t\|\g^{(I)}_j(s)\|^2ds, \\
&\s_{m,n}^2(T_m-t)=\int^{T_m}_t\left\|\sum_{i=m+1}^n \left[\a_i(t)
\g^{(I)}_i(s)+(\a_i(t)-w_i(t))\Sigma_i(s)\right]\right\|^2ds,
\end{split}\label{e742}
\end{equation}
for some index $k$, $j$, and pairs of indexes $m$ and $n$ in the input set.

We can take either a parametric or a non-parametric approach for calibration. In the non-parametric approach,
the volatilities of inflation rates, $\g^{(I)}_j(t)$, are assumed piece-wise functions of $t$. The number of unknowns is usually big and thus equations (\ref{e742}) will often be under-determined and thus ill-posed. Regularization
is usually needed in order to achieve uniqueness and smoothness of solution. An efficient technique
is to impose a
quadratic objective function for both uniqueness and smoothness (Wu, 2003). When both objective function
and constraints, listed in (\ref{e742}), are quadratic functions, the constrained optimization
problem can be solved with a Hessian-based descending search algorithm, where each step of iterations only requires
solving a symmetric eigenvalue problem, and is thus very efficient. For the details of such a methodology, we refer to
Wu (2003).

For demonstrations, we consider calibrating a two-factor model where the inflation rates are driven by one factor while the nominal rates are driven by another factor.
Let $\rho$ be the correlation between the nominal rate and inflation rate, then
(\ref{e742}) becomes,
\begin{equation}
\begin{split}
&\s_j^2(T_j-t)=\int^{T_j}_t|\g^{(I)}_j(s)|^2ds, \\
&\s_{m,n}^2(T_m-t)=\int^{T_m}_t\sum_{i,j=m+1}^n \left[ \a_i(t)\a_j(t)\g^{(I)}_i(s)\g^{(I)}_j(s)\right . \\
&\qquad\qquad\qquad\qquad +2\a_i(t)(\a_j(t)-w_j(t))\g^{(I)}_i(s)\Sg_j(s) \rho\\
&\qquad\qquad\qquad\qquad \left.+(\a_i(t)-w_i(t))(\a_j(t)-w_j(t))\Sg_i(s)\Sg_j(s)\right]ds,
\end{split}\label{e743}
\end{equation}
where $\g^{(I)}(s)$ are scalar functions, and $\Sigma_i(s)$ is a known function such that
\[
\Sigma_i(s)=-\sum^i_{l=\eta_t}{\D T_{l+1}f_l(t)\over 1+\D T_{l+1}f_l(t)}\g_l(s).\]

If we take the approach of non-parametric calibration by assuming piece-wise constant function for $\g^{(I)}_j$,
we then have a set of linear or quadratic functions to solve. By adding a quadratic objective function, say,
\[
O(\{\g^{(I)}_j\})=\a\sum (\g^{(I)}_j-\g^{(I)}_{j-1})^2,\]
we make the problem well-posed and easy to solve numerically. Here $\a>0$ is a weight parameter.

We can also back out the implied correlation. To do so, we may assume piece-wise correlation,
$\rho(t)=\rho_i$ for $T_{i-1}\leq t<T_i$, and use instead the following objective function:
\begin{equation}\label{e541}
O(\{\g^{(I)}_j\})=\a\sum (\g^{(I)}_j-\g^{(I)}_{j-1})^2+\b\sum(\rho_i-\rho_{i-1})^2, \quad \a>0,\b>0.
\end{equation}
In addition, we need to impose $-1\leq \rho_i\leq 1$. Given that both the objective function (\ref{e541}) and
constraints (\ref{e743}) are quadratic functions, the method developed by Wu (2003) should work well.

As an example, we calibrate the two-factor market model to price data of Euro ZCIIS and
inflation caps as of April 7, 2008\footnote{We do not have the data of YYIIS or swaptions.}, tabulated
in Table 1 and 2, respectively. The payment frequency for both types of instruments is annual (i.e. $\D T_j=\D T=1$), and the cap prices are given in basis points (bps).
The input correlation between the nominal
and the inflation rates is estimated using data of the last three years,
from January 2005 to February 2008, and the numbers is $\rho=-5.35\%$.
For simplicity we have taken a flat volatility for all nominal forward rates, at the level of 15\%. The calibration also makes use of
the LIBOR data, including LIBOR rates, swap rates and prices of at-the-money (ATM)
caps, which are not included in the paper for brevity\footnote{The data are available upon request; or one can find them in Bloomberg.}.
\begin{center}
Table 1. Swap rates for ZCIIS for 2008/4/7 \\
\begin{tabular}{|c|c|} \hline
Maturity (Year) & Swap Rate (\%) \\ \hline\hline
1 & 2.2115	\\ \hline 	
3 & 2.3920	\\ \hline	
5 & 2.3500	\\ \hline	
7 & 2.3425	\\ \hline	
10 & 2.3530	\\ \hline	
15 & 2.3830	\\ \hline	
20 & 2.3870	\\ \hline	
25 & 2.4065	\\ \hline	
30 & 2.4315 \\ \hline \hline
\end{tabular}
\end{center}

\bigskip
\begin{center}
Table 2. Prices (in bps) of inflation caps in 2008/4/7 \\
\begin{tabular}{|c|r|r|r|} \hline
 & \multicolumn{3}{c|}{Strike (\%)} \\ \hline
Mat. & 2 & 3 & 4 \\ \hline\hline
2 & 101.6		&	21.1		&	3.5  \\ \hline 	
3 & 157.7		&	30.9		&	7.7 \\ \hline	
5 & 253.2		&	62.3		&	13.5 \\ \hline	
7 & 349.1		&	93.7		&	21.3 \\ \hline	
10 & 491.6		&	143.7		&	37.7 \\ \hline
12 & 582.3		&	179.3		&	50.8	    \\ \hline
15 & 709.9		&	230.2		&	73.4 	\\ \hline	
20 & 911.8		&	326.2		&	121.8	\\ \hline		
30 & 1229.7		&	494.4		&	216.4
 \\ \hline \hline
\end{tabular}
\end{center}

We first construct the term structure of inflation rates, using nominal
and inflation discount factors. The term structure is displayed in
Figure 2, together with the term structure of nominal forward rates. One can see that the magnitude of the inflation forward rates is consistent with that of ZCIIS rates, and the two curves show a low degree of negative correlation.

\bigskip

\centerline{\epsfxsize=4in \epsfbox{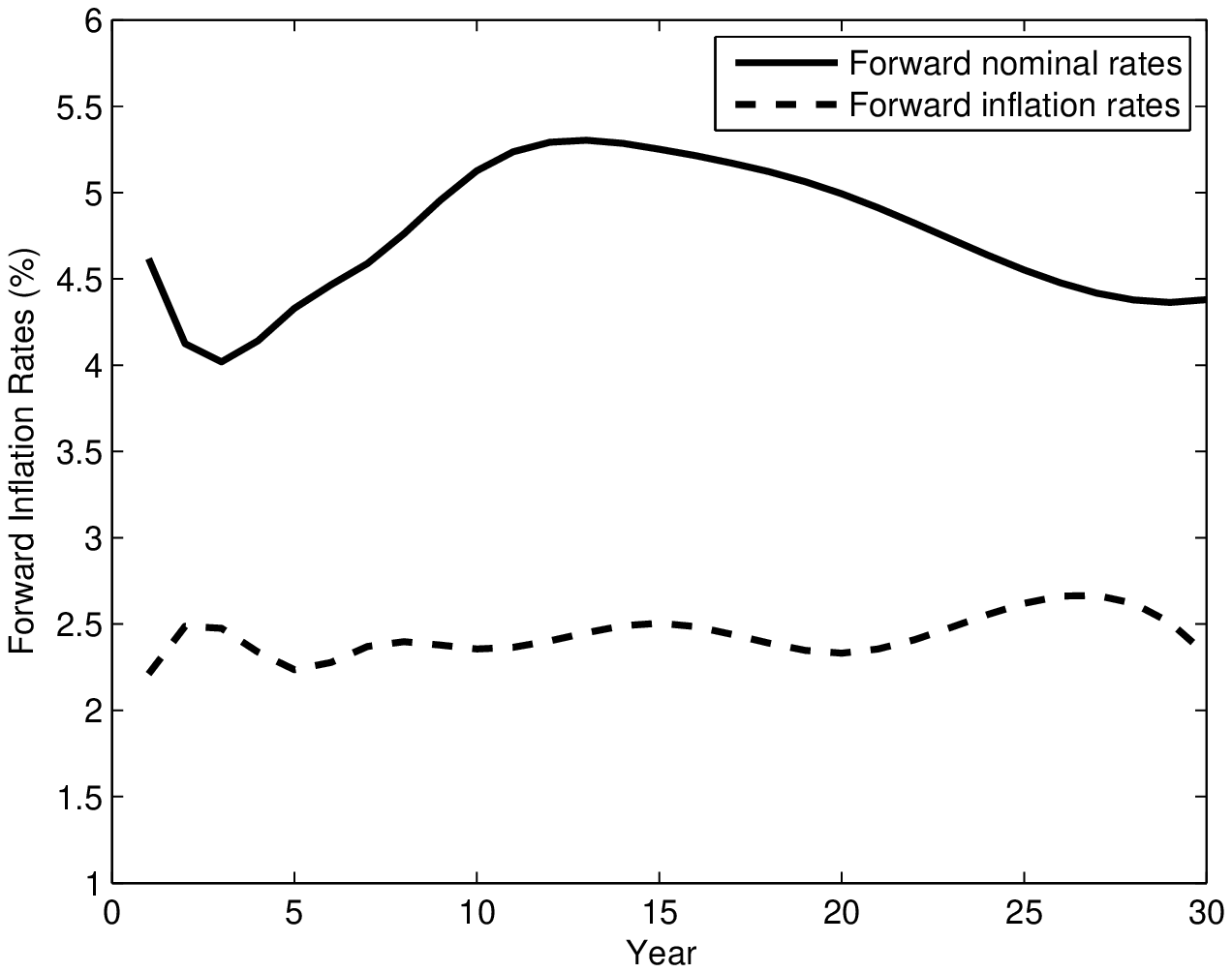}}
\centerline{{\bf Figure 2}  Term structure of the nominal forward rates }
\centerline{and inflation forward rates.}

\bigskip

We then proceed to backing out the implied volatilities of the displaced inflation forward rates, $\s_j$'s, and set  $\g^{(I)}_j(t)=\s_j, \forall t\leq T_j$.
The procedure consists of two steps. First we need to bootstrap the caplet prices, then we solve for $\s_j$'s
through a root-finding procedure using formulae (\ref{MM5}) and (\ref{e537}).
Note that in its current form the market cannot price volatility smiles or skews\footnote{To calibrate to more strikes we will need a smile model.}, so we have only tried to
calibrate to caps for strike $K=2\%$.
The results are displayed in Figure 3. One can see that the local volatility
varies around 0.5\%, which is the magnitude of implied volatilities often observed in the market.

\bigskip

\centerline{\epsfxsize=4in \epsfbox{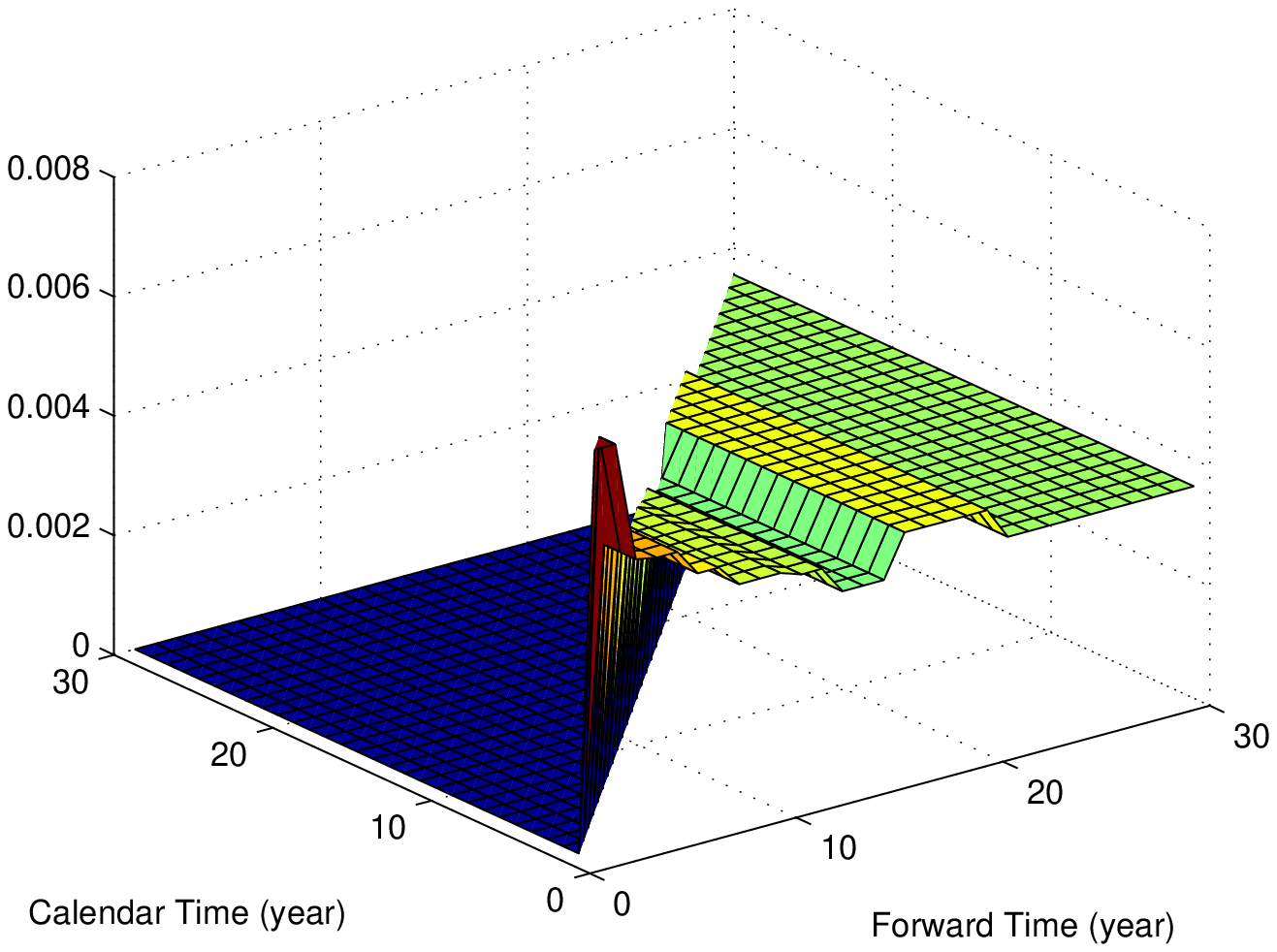}}
\centerline{{\bf Figure 3}  Calibrated local volatility surface, $\g^{(I)}_i(t)$.}



\bigskip

Next, we price inflation swaptions using the calibrated model. The spot swap-rate
curve is displayed in Figure 4, which is also slightly upward sloping.

\bigskip

\centerline{\epsfxsize=4in \epsfbox{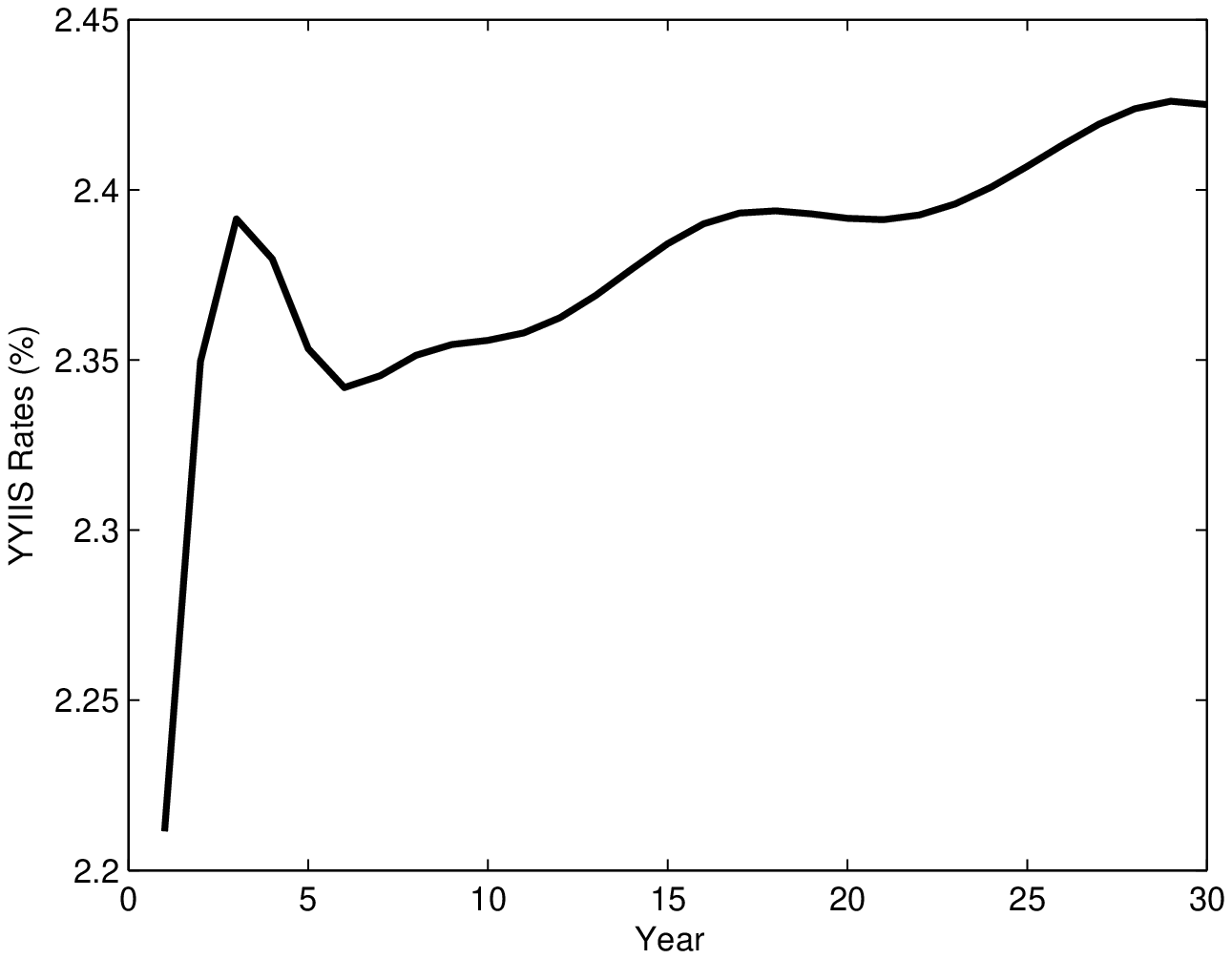}}
\centerline{{\bf Figure 4}  Term structure of the inflation swap
rates.}

\bigskip

For various maturities, tenors and strikes, we calculate prices of inflation swaption by (\ref{IIS16}).
The results are presented in dollar prices in Figure 5 - 8. One can see that the prices
vary in a reasonable and
robust way according to maturities, tenors and strikes.

\bigskip

\centerline{\epsfxsize=4in \epsfbox{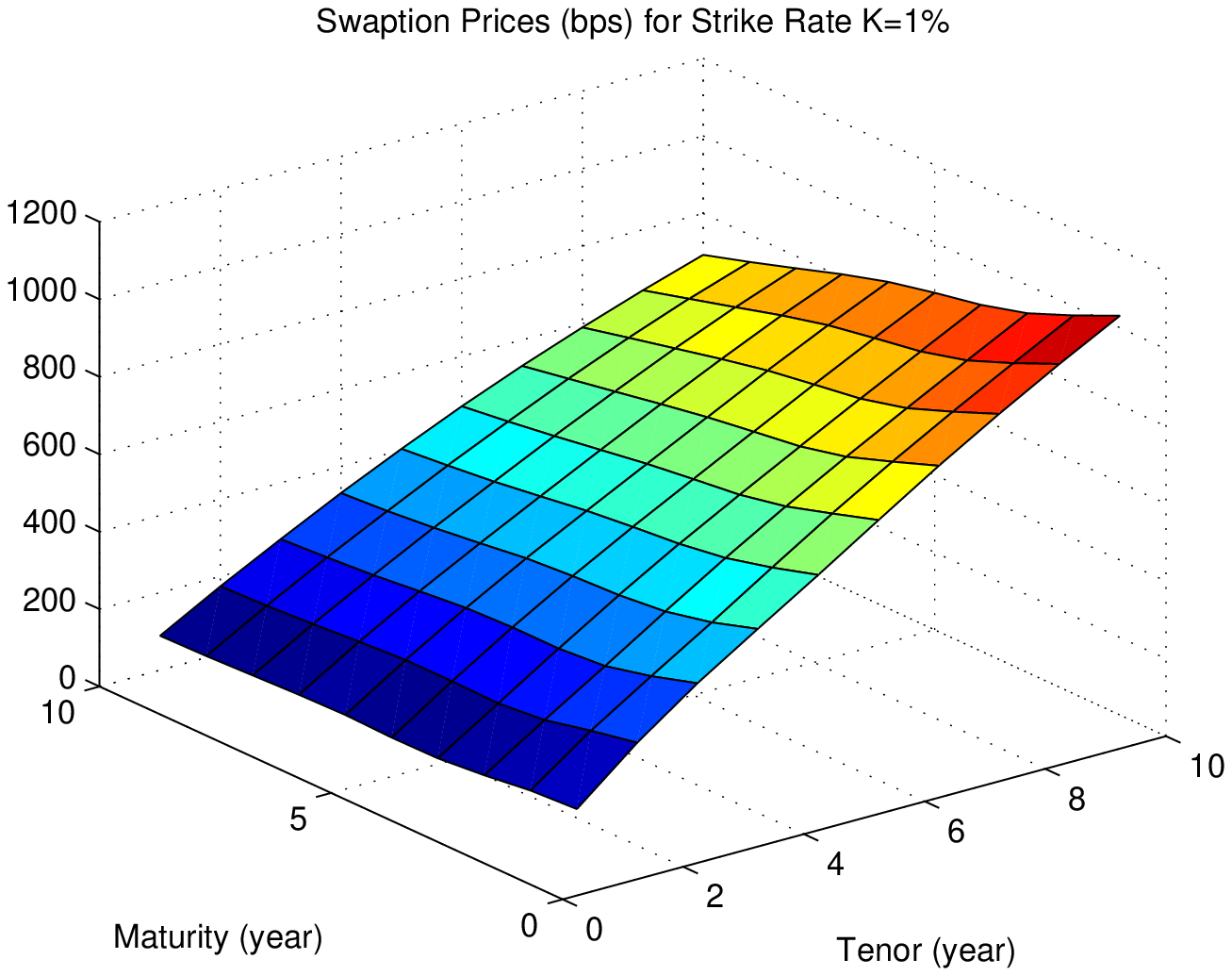}}
\centerline{{\bf Figure 5} Price surface of swaptions for $K=1\%$.}

\bigskip

\centerline{\epsfxsize=4in \epsfbox{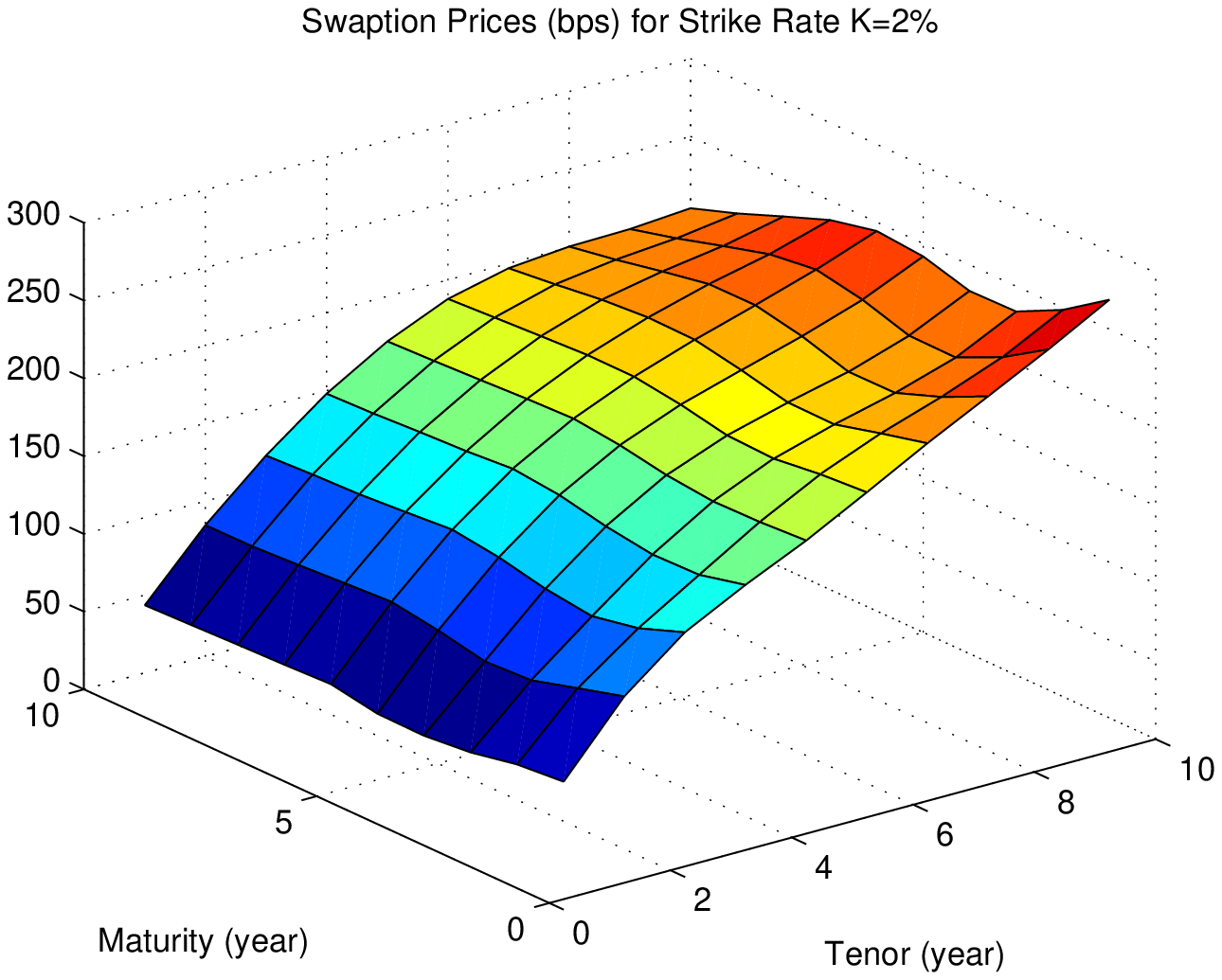}}
\centerline{{\bf Figure 6} Price surface of swaptions for $K=2\%$.}

\bigskip

\centerline{\epsfxsize=4in \epsfbox{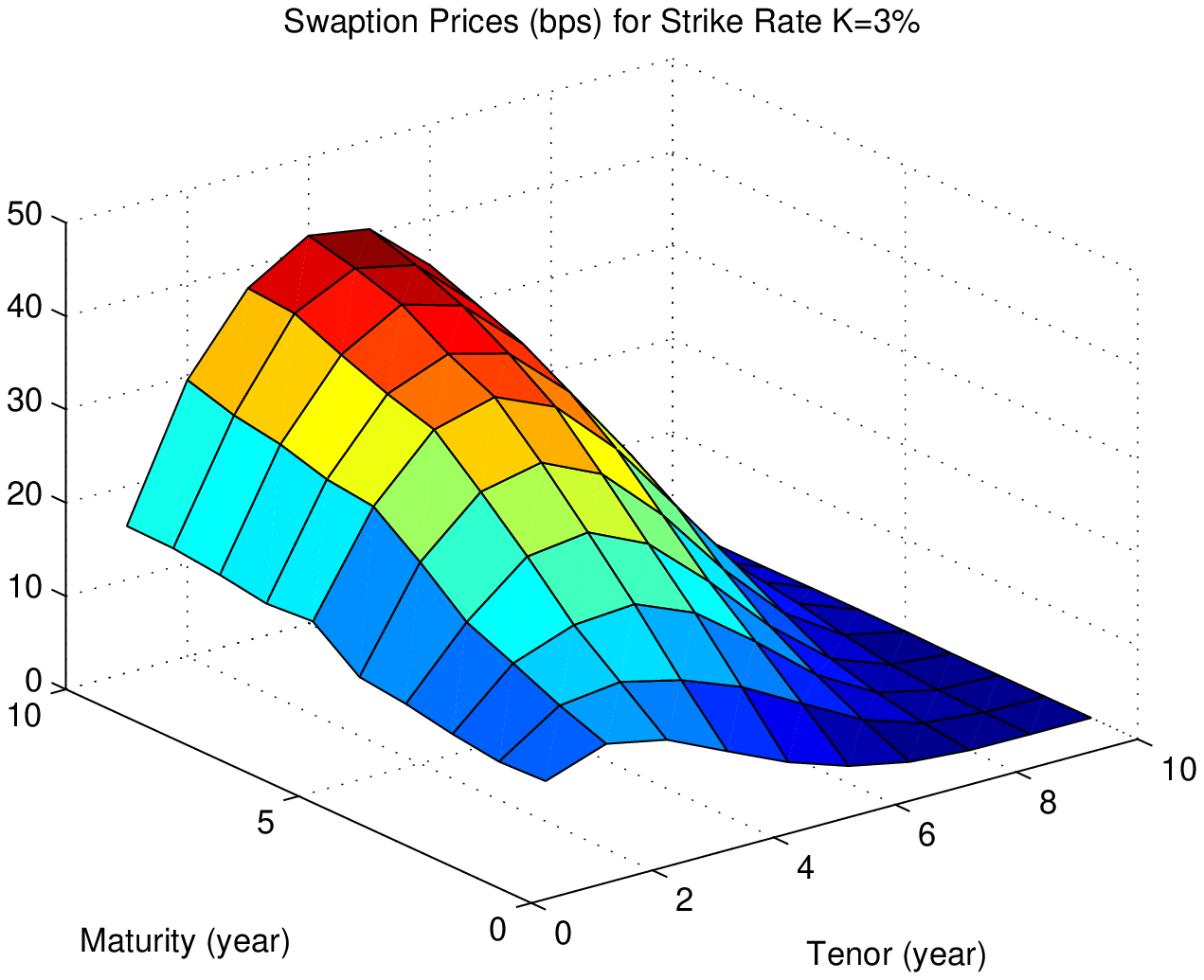}}
\centerline{{\bf Figure 7} Price surface of swaptions for $K=3\%$.}

\bigskip

\centerline{\epsfxsize=4in \epsfbox{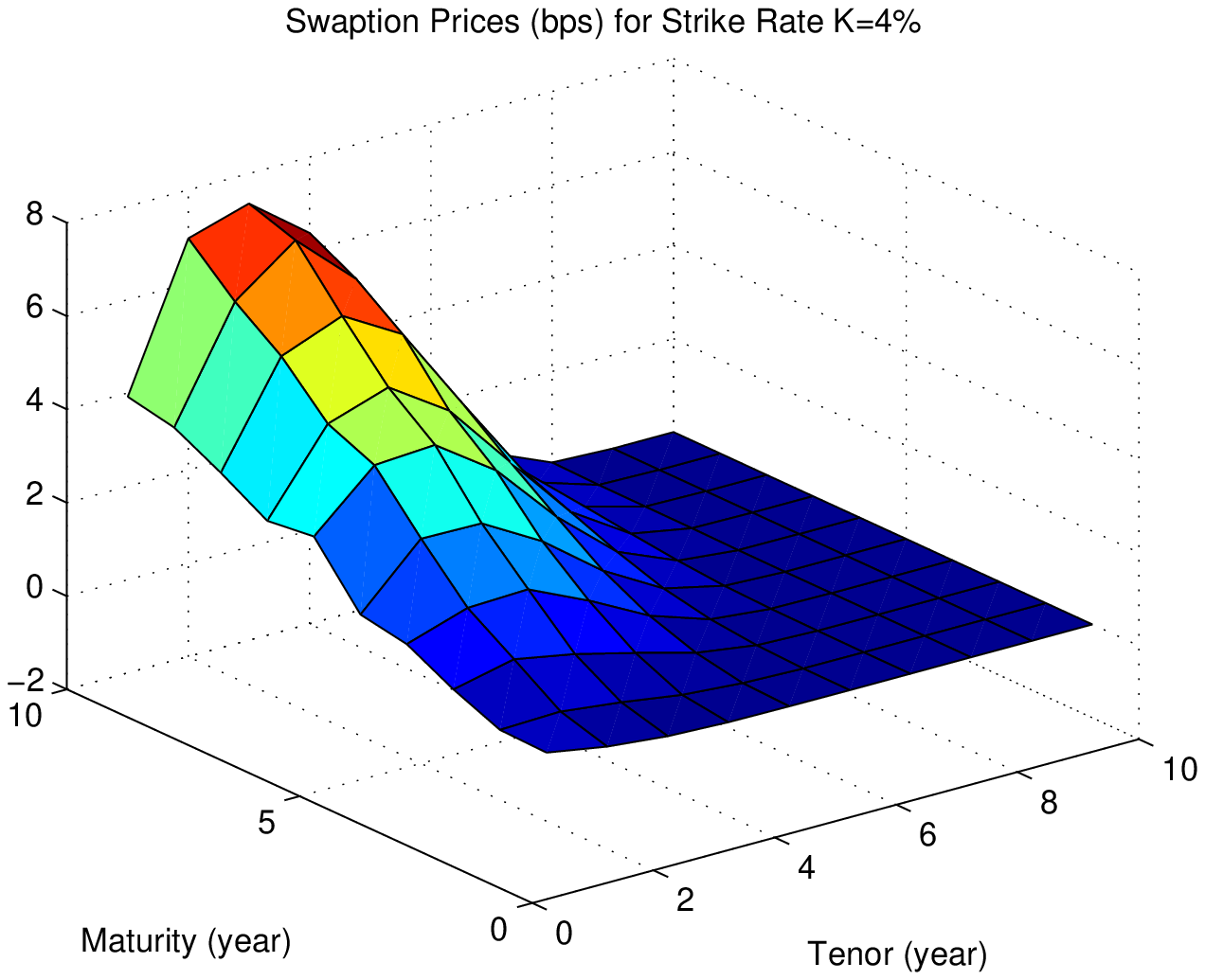}}
\centerline{{\bf Figure 8} Price surface of swaptions for $K=4\%$.}

\bigskip






\section{Smile Modeling Based on the Market Model}

It is well known that inflation caps and floors demonstrate so-called the implied volatility smiles. Having developed the market models, we can proceed to cope with volatility smiles in ways similar to smile modeling for interest-rate derivatives based on LIBOR market model, which, routinely, involve with adopting additional risk factors like stochastic volatilities or jumps, or taking level-dependent volatilities. For example, we may adopt the SABR dynamics for the expected displaced inflation forward rates, $\mu_i(t)$, and develop the following model:
\begin{equation}
\left\{
\begin{split}\label{e847}
d\mu_j(t)&=\mu_j^{\b_j}(t)\a_j(t) d{Z}_t^j, \\
d\a_j(t)&=\nu_j\a_j(t)dW_t^j,
\end{split}\right .
\end{equation}
where $\b_j$ and $\nu_j$ are constants, both ${Z}_t^j$ and $W_t^j$ are one-dimensional Brownian motions under the $T_j$-forward measure, which can be correlated,
\[
dZ_t^jdW_t^j=\rho_jdt.\]
Mecurio and Mereni (2009) proposed and studied the above model with $\b_j=1$, and demonstrate a very quality fitting of implied volatility smiles with the model.

We can also consider other extensions of the market model for smile modeling yet, given the rich literature on smile modeling of interest-rate derivatives, the extensions may become some sort of routine exercises. We refer readers to Brigo and Mercurio (2006) and for an introduction of major smile models for interest-rate derivatives based on the LIBOR market model. Of course, empirical study with various smile models for inflation rates should be an interesting as well as challenging issue.

\section{Conclusion}
Using prices of real zero-coupon bonds as model primitives that are tradable through ZCIIS, we define the term structure of inflation rates, and then construct a market model as well as a HJM type model for the term structure of inflation rates. We show that the HJM type model with inflation forward rates is consistent with the HJM model with real forward rates developed through ``foreign currency analogy". The market can be used to price inflation caplets/floorlets and swaptions in closed form, and can be calibrated efficiently. Finally, the current model serves as a platform for further extensions using risk dynamics in addition to diffusions.

\appendix
\section{Proofs of Propositions}
\noindent \emph{Proof of Proposition 1}:

Do the following zero-net transactions.
\begin{enumerate}
\item At time $t$,
\begin{enumerate}
\item Long the forward contract to buy ${I(T_1)\over I(T_0)}$ dollar worth of $T_2$-maturity real bond deliverable at $T_1$ at the unit price $F_R(t,T_1,T_2)$ (i.e., to buy ${I(T_1)\over I(T_0)F_R(t,T_1,T_2)}$ units);
\item long one unit of $T_1$-maturity real bond at the price of $P_R(t,T_0,T_1)$;
\item short ${P_R(t,T_0,T_1)\over P_R(t,T_0,T_2)}$ unit(s) of $T_2$-maturity real bond at the price of $P_R(t,T_0,T_2)$.
\end{enumerate}
\item At time $T_1$, exercise the forward contract to buy the $T_2$-maturity real bond (that pays $I(T_2)/I(T_1)$) at the price $F_R(t,T_1,T_2)$, applying all proceed from the $T_1$-maturity real bond.
\item At Time $T_2$, settle all transactions.
\end{enumerate}
The profit or loss from the transactions is
\begin{equation}
P\&L=\left({1\over F_R(t,T_1,T_2)}-{P_R(t,T_0,T_1)\over P_R(t,T_0,T_2)}\right){I(T_2)\over I(T_0)}.
\end{equation}
For the absense of arbitrage, the forward price must be equal to (\ref{e531}) $\Box$
\\

\noindent \emph{Proof of Proposition 5}

According to (\ref{e640}),
\begin{equation}\label{IIS4}
S_{m,n}(t)+\frac{1}{\Delta T_{m,n}} = \sum^n_{i=m+1}\o_i(t)\mu_i(t),
\end{equation}
so the dynamics of the displaced swap rate will arise from, by Ito's lemma,
\begin{equation}\label{IIS5}
\begin{split}
d\left(S_{m,n}(t)+\frac{1}{\Delta T_{m,n}}\right)&=
\sum^n_{i=m+1}\mu_i(t)d\o_i(t)+\o_i(t) d\mu_i(t)+d\o_i(t) d\mu_i(t).
\end{split}
\end{equation}
One can easily show that
\begin{equation}\label{a056}
d\o_i(t)=\o_i(t)(\Sg_i(t)-\Sg_A(t))\cdot(d{\bf Z}_t-\Sg_A(t)dt),
\end{equation}
where $\Sg_A(t)=\sum^n_{i=m+1}\o_i\Sg_i(t)$.
Making use of (\ref{e534}) and (\ref{a056}), we have
\begin{equation*}
\begin{split}
&d\left(\sum^n_{i=m+1}\o_i(t)\mu_i(t)\right)=\sum^n_{i=m+1}\o_i(t)\mu_i(t)\left[(\Sg_i(t)-\Sg_A(t))\cdot(d{\bf Z}_t-\Sg_A(t)dt) \right. \\
&\qquad\qquad\qquad \left .+\g^{(I)}_i(t)\cdot(d{\bf Z}_t-\Sg_i(t)dt) +\g^{(I)}_i(t)\cdot(\Sg_i(t)-\Sg_A(t))dt \right] \\
=&\sum^n_{i=m+1}\o_i(t)\mu_i(t)\left(\Sg_i(t)-\Sg_A(t)+\g^{(I)}_i(t)\right)\cdot(d{\bf Z}_t-\Sg_A(t)dt)\\
=&\left(\sum^n_{i=m+1}\o_i(t)\mu_i(t)\right) \\
&\qquad\qquad\times\left[\sum^n_{i=m+1}\a_i(t)\left(\g^{(I)}_i(t)+\Sg_i(t)\right)-\Sg_A(t)\right]\cdot(d{\bf Z}_t-\Sg_A(t)dt)
\end{split}
\end{equation*}
which is (\ref{e642}).

Finally, we point out that $d{\bf Z}_t-\Sg_A(t)dt$ is a Brownian motion under the martingale measure corresponding to
the numeraire $A_{m,n}(t)$. Let $Q_{m,n}$ denote this measure, then it is defined by
the Radon-Nikodym derivative with the risk neutral measure by $Q$
\begin{equation*}\label{IIS11}
\left. \frac{dQ_{m,n}}{dQ}\right|_{\calF_t} =\frac{A_{m,n}(t)}{A_{m,n}(0)B(t)}= m_s(t) \quad \mbox{for} \quad t \leq T_n,
\end{equation*}
where $B(t)$ be the money market account under discrete compounding:
\begin{equation*}
B(t)=\left(\prod^{\eta_t-2}_{j=0}(1+f_j(T_j)\D T_j)\right) \left(1+f_{\eta_t-1}(T_{\eta_t-1})(t-T_{\eta_t-1})\right).
\end{equation*}
By Ito's lemma,
\begin{equation}\label{IIS12}
dm_s(t) = m_s(t) \Sigma_A(t)\cdot \dbz_t.
\end{equation}
The $Q_{m,n}$ Brownian motion corresponding to ${\mbox{\boldmath
$Z$}}_t$ is defined by
\begin{eqnarray}\label{IIS13}
\dbz^{(m,n)}_t &=& \dbz_t - \left<\dbz_t,
\frac{dm_s(t)}{m_s(t)}\right> \nonumber \\
&=& \dbz_t - \Sigma_A(t) dt\quad \Box
\end{eqnarray}

\end{document}